\documentclass[aps,twocolumn,showpacs,superscriptaddress,groupedaddress,amssymb,amsfonts,amsmath,letterpaper]{revtex4-2}  

\usepackage{graphicx}
\usepackage{amsmath}
\usepackage{upgreek}
\usepackage{cancel}
\usepackage{natbib}
\usepackage{color}
\usepackage{hyperref}
\definecolor{darkblue}{rgb}{0.0,0.0,0.3}
\hypersetup{colorlinks,breaklinks,
            linkcolor=darkblue,urlcolor=darkblue,
            anchorcolor=darkblue,citecolor=darkblue}

\newcommand{\mnras}{Mon.~Not.~Roy.~Astron.~Soc.}

\renewcommand{\apj}{Astrophys.~J.}

\renewcommand{\prl}{Phys.~Rev.~Lett.}

\renewcommand{\pre}{Phys.~Rev.~E}

\newcommand{\araa}{Ann.~Rev.~Astron.~Astro.}

\newcommand{\D}[2]{\frac{{\rm d} #2}{{\rm d} #1}}
\newcommand\bb[1]{\mbox{\boldmath{$#1$}}}

\newcommand{\rmd}{{\rm d}}

\begin{document}

\title{Amplification of turbulence through multiple planar shocks}

\author{Michael F.~Zhang}\email{mfzhang@princeton.edu}
\affiliation{Department of Astrophysical Sciences, 
Princeton University, Peyton Hall, Princeton, NJ 08544, USA}
\affiliation{Princeton Plasma Physics Laboratory, PO~Box 451, Princeton, NJ 08543, USA}
\author{Seth Davidovits}
\affiliation{Lawrence Livermore National Laboratory, Livermore, California 94550, USA}
\author{Nathaniel J.~Fisch}
\affiliation{Department of Astrophysical Sciences, 
Princeton University, Peyton Hall, Princeton, NJ 08544, USA}
\affiliation{Princeton Plasma Physics Laboratory, PO~Box 451, Princeton, NJ 08543, USA}

\date{\today}

\begin{abstract}
	We study the amplification of isotropic, incompressible turbulence through multiple planar, collisional shocks, using analytical linear theory. There are two limiting cases we explore. The first assumes shocks occur rapidly in time such that the turbulence does not evolve between shocks. Whereas the second case allows enough time for turbulence to isotropize between each shock. For the latter case, through a quasi-equation-of-state, we show that the weak multi-shock limit is agnostic to the distinction between thermal and vortical turbulent pressures, like an isotropic volumetric compression. When turbulence does not return to isotropy between shocks, the generated anisotropy -- itself a function of shock strength -- can feedback on amplification by further shocks, altering choices for maximal or minimal amplification. In addition for this case, we find that amplification is sensitive to the shock ordering. We map how choices of shock strength can impact these amplification differences due to ordering, finding, for example, shock pairs which lead to identical mean post-shock fields (density, temperature, pressure) but maximally distinct turbulent amplification.

\end{abstract}

\maketitle

\section{Introduction}
\label{sec:introduction}

The interaction of turbulence with shocks is a fundamental fluid phenomenon that is ubiquitous across a variety of fields. For example, turbulence seeded by asymmetry or Rayleigh-Taylor and Richtmeyer-Meshkov instabilities in inertial confinement fusion (ICF) implosions \cite{veli2007,hammel2010,weber2014,ma2013,zhou2024} where shocks are used in compression and are present during stagnation \cite{lindl1995}; accretion shocks e.g. supernovae explosions \cite{hillebrandt2000,blondin2003,abdikamalov2016} and star formation \cite{maclow2004}; supersonic flight and propulsion \cite{tam1982}.

Often, there may be multiple shocks in physical situations of interest where the interaction with turbulence or fluctuations can be important. In ICF, multiple shocks are used for the efficient compression of the fuel capsule \cite{lindl1995,boehly2011}. The interaction of the first shock with density non-uniformity in the target, such as grains \cite{davidovits2022a,davidovits2022b,li2024} or in hydrocarbon foams filled with deuterium-tritium \cite{sacks1987,moody2000,root2013,kotelnikov1998}, can generate a turbulent flow field that interacts with subsequent shocks \cite{huete2012}. In addition, some experiments at the National Ignition Facility (NIF)  are investigating the Richtmeyer-Meshkov instability (RMI) and can feature multiple planar shocks \cite{nagel2017,nagel2022}. Related to this, there have been computational studies to understand how RMI behaves under the action of multiple shocks \cite{bender2021,leinov2009}. Within astrophysics, turbulence is important in understanding the structure of interstellar gas and molecular clouds, where multiple shocks may be present \cite{pudritz2013,kupilas2021}. In supersonic propulsion, trains of shocks manifest within nozzles, and can interact with turbulent boundary layers \cite{grzona2011}.
In many of these cases, the conditions under which turbulence would be maximally or minimally enhanced by a sequence of multiple (or many) shocks is important to the dynamics. Here we will use a theory for shock-turbulence interaction to study this and other dynamics for multiple planar shocks.

Many works thus far have been focused on the canonical interaction of a single planar shock with isotropic turbulence \cite{ribner1969,ribner1987,ribner1998}. Fundamental theoretical studies were initiated by Ribner's Linear Interaction Analysis (LIA) \cite{ribner1953,ribner1954}, which took a Kovasnay decomposition \cite{kovasnay1953} of the turbulence into a superposition of shear wave modes. Through linearizing the fluid equations and Rankine-Hugoniot jump relations under the assumption of weak turbulence relative to the shock, how each mode changes across the shock can be determined. The effects can then be summed to infer properties of the downstream turbulence, including the amplification of turbulent kinetic energy.

Later works have revisited LIA and developed analytical expressions for amplification of turbulent kinetic energy in general and in certain asymptotic limits \cite{wouchuk2009}. Others have made use of Moore's results on the linear interaction of an acoustic wave with a shock \cite{moore1953} to adapt the analysis for upstream acoustic and entropy modes \cite{mahesh1995,mahesh1997}.

In this paper, we adapt Ribner's single-shock LIA to develop an initial theory for the multi-shock amplification of initially isotropic (or axisymmetric), incompressible, vortical, weak turbulence. For such turbulence, good agreement on turbulent kinetic energy (TKE) amplification has been found between Ribner's single-shock LIA with direct numerical simulation (DNS) \cite{jamme2002,ryu2014,chen2019} and a wind tunnel experiment \cite{barre1996,ribner1998}. Whilst there have been discrepancies on the downstream anisotropy \cite{larsson2009} and when departing from the assumptions of LIA \cite{lee1993,lee1997,larsson2013,mahesh1994,hannappel1994}, the otherwise good agreement motivates the use of LIA as a building block for a theory of multi-shock interaction with turbulence. We do this for two opposite limits where either the turbulence is allowed to isotropize between shocks, or it remains anisotropic. Taking such limits allows for the use of inviscid LIA to compute turbulent vortical spectra without concern for the details of the nonlinear, viscous evolution between each shock. The isotropy, or not, of turbulence under compression has been shown to be important for its growth in the case of uniform (metric) compression \cite{davidovits2020}, and so we examine both limits here.

Our model is described further in Section~\ref{sec:model}. We give formulae that prescribe the TKE amplification in each limit. In particular, amplifications of longitudinal (parallel to shock propagation) and lateral (perpendicular to shock propagation, parallel to shock front) components of vortical turbulence are given in the non-isotropized limit.

We compute these as functions of shock strength for various multi-shock compression scenarios of interest in Section~\ref{sec:results}. General trends with shock strength are shown in each limit for a series of identical shocks, in Section~\ref{sec:trends}. We show that weaker shocks are more efficient at amplifying turbulence for a fixed final density state. In particular, for the isotropized limit, there exists an optimal shock strength that depends only on the thermal polytropic index. In the non-isotropized limit, we will show that the preferential longitudinal enhancement of turbulence by weaker shocks sustains the continued amplification by subsequent shocks. However, a series of stronger shocks preferentially amplifies the lateral component through a refractive effect and can simultaneously suppress the longitudinal component, limiting further amplification.

We use these calculations, in Section~\ref{sec:eos}, to inform compression energetics by acquiring polytropic indices via a turbulent quasi-equation-of-state. These indices show that the turbulent pressure in the weak-shock, isotropized limit exhibits the same adiabatic behavior as the thermal pressure for an isotropic, metric compression. In contrast, the non-isotropized case is super-adiabatic in the weak shock limit. Departing from the weak shock limit reduces the polytropic indices associated with turbulent pressure in both limits, leading to sub-adiabatic behavior under shock compression.

One significant departure between the two limits is discovered for a series of non-identical shocks of variable strength, in Section~\ref{sec:ordering} . We show such a scenario is non-commutative for the non-isotropized limit, whereby the amplification of TKE is sensitive to the order of the shocks. In Section~\ref{sec:discussion}, we discuss the implications of these results for experimental shock-compression scenarios, and the caveats to the assumptions made in the model.

\section{Model}
\label{sec:model}

We now discuss our model for the multi-shock amplification of turbulence, and the simplifying assumptions used. Consider low-Mach, isotropic, homogeneous, incompressible turbulence in a compressible flow. Assuming there are no large macroscopic gradients, the dynamics are well described by the compressible Navier-Stokes equations, and one could assume an ideal gas equation of state.

The parameters that describe the flow are: density $\rho$, flow velocity $\mathbf{v}$, and thermal pressure $p$, which relates to the total energy, $E_{\rm total}$, via the equation of state. We assume $\gamma=5/3$ for the thermal polytropic index. The three spatial dimensions are $x$, $y$ and $z$, with the shock normal aligned to $x$. $\mathbf{v}= \bar{\mathbf{v}} + \tilde{\mathbf{v}}$  includes the background flow, $\bar{\mathbf{v}}$, and the turbulent flow, $\tilde{\mathbf{v}}$. For brevity, we will denote the turbulent kinetic portion of the energy as 
\begin{equation} \label{eq:e_def}
	E \doteq E_{\rm TKE} \doteq \int\rmd^3\bb{r} \frac{1}{2}  \rho \tilde{\mathbf{v}}^{2} =  \frac{1}{2}  \rho V \overline{\tilde{\mathbf{v}}^{2}},
\end{equation}
where we have used the incompressibility and homogeneity of the turbulence, assuming acoustic modes are negligible, and the bar denotes a volume average.

With appropriate initial and boundary conditions, a shock can propagate past the turbulence. Both the mean properties of the flow and the root mean square amplitude of the turbulent fluctuations will change across the shock. The strength of the shock measures  how large the jumps in fluid quantities are. It can be described either by the Mach speed of the shock relative to the background flow, or any of the jumps in mean fluid properties such as the density, $m = \rho_{1}/{\rho_{0}}$, where the subscript refers to the background quantity after that number of shocks.

\begin{figure}
    \centering
    \includegraphics[width=0.8\columnwidth]{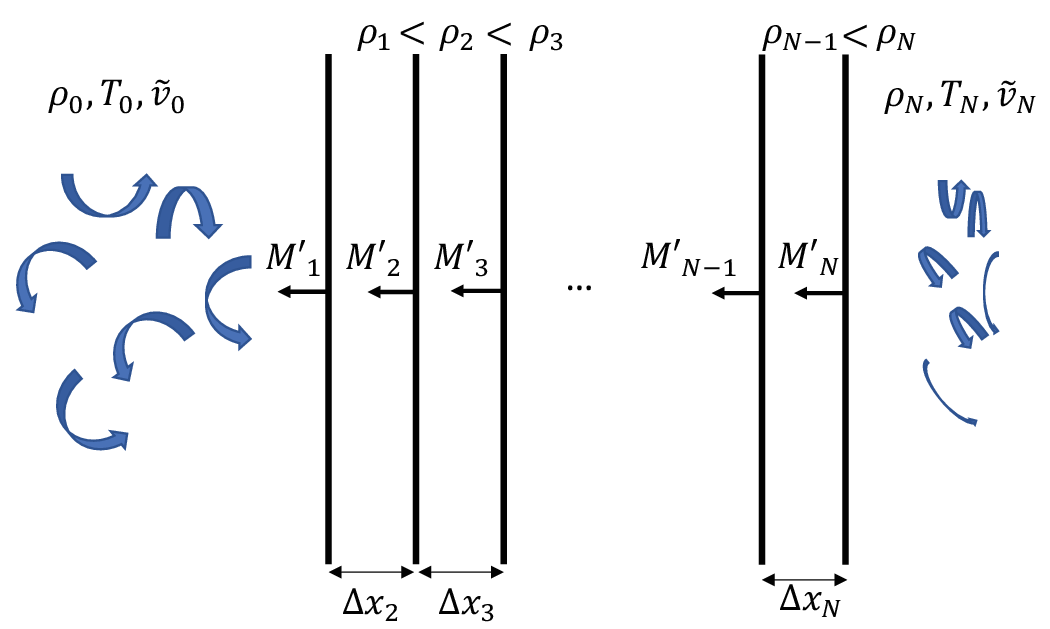}
    \caption{Setup of problem for interaction of an initial isotropic distribution of incompressible turbulence on the left with N leftward propagating shocks. The Mach speeds of each shock in the lab frame, denoted by a prime, $M'_{i} $, are not the same as those used in calculating shock-jump relations, $M_{i}$, where the background flow ahead of a shock is taken to be stationary. Fluid quantities downstream of each shock are denoted by the number of the shock, with $0$ denoting the initial upstream conditions. The spacing between shocks, $\Delta x_{i}$, can be arbitrarily adjusted such that they do not overtake each other before propagating through the turbulence.}
    \label{fig:setup}
\end{figure}

A simple setup for investigating the amplification of isotropic, incompressible turbulence after multiple planar shocks is shown in Fig. \ref{fig:setup}. We imagine choosing N shocks of given strengths, defined by their density jumps, 
\begin{equation} \label{eq:m_def}
	m_{i} \doteq \frac{\rho_{i}}{\rho_{i-1}},
\end{equation}
for shock $i$. The shocks are spaced such that the following shocks do not overtake the preceding shocks before propagating past the turbulent region. These density jumps, $m_{i}$, give a frame-independent measure of shock strength, whereas the Mach speeds in the lab frame, $M'_{i}$, to achieve a given $m_{i}$ generally depend on the background flow speed immediately ahead of a shock.

Two cases can then be examined in this problem: a) Spacing between shocks is short, such that the turbulence does not have sufficient time to evolve between shocks. For a turbulent turnover time, $\tau_{\epsilon} \sim  E/\D{t}{E}$ \cite{sagaut2018}, this condition can be expressed as ($\Delta t_{i} = \frac{\Delta x_{i}}{M'_{i}}   \ll \tau_{\epsilon}$). We will refer to this as the non-isotropization case (NIC). b) The second shock follows a sufficient distance behind the first shock such that the anisotropic, shocked turbulence returns to isotropy before entering the second shock, and similarly for proceeding shocks. This latter limit will be denoted as the isotropization case (IC).

In both cases, by assuming sufficiently low-Mach (weak) turbulence, the overall jumps in average background quantities will be approximately unperturbed \cite{chen2019}. Therefore, the overall jump after $N$ shocks can be calculated as the product from each of the single-shock jump relations, given by Rankine-Hugoniot. However, the overall jump of turbulent velocity will depend on whether it returns to isotropy between shocks or not. As such, the two limiting cases will differ.

An idealized, simplified treatment of the IC limit can be achieved by neglecting viscous dissipation during the return to isotropy (this is idealized because the turbulence will experience some decay in energy during the return to isotropy \cite{choi2001}). This is a convenient limit to draw direct comparison to turbulent amplification under an isotropic, uniform (metric) compression. Under the assumption of no viscous dissipation, the overall jump in turbulent amplitude can be calculated in a similar manner to the background quantities, by evaluating the product of the single-shock jumps. For the turbulent velocity, the single-shock jump in amplitude can be calculated using results from Linear Interaction Analysis (LIA), given in Ribner \cite{ribner1954}. LIA has been found to agree with direct numerical simulations (DNS) for low mach turbulence \cite{ryu2014}. The final turbulent amplification after $N$ shocks  can then be calculated from the single-shock jumps in a similar manner to the background quantities,

\begin{equation} \label{eq:gen_amp}
	\frac{A_{N}}{A_{0}} = \prod_{i=1}^{N}  \frac{A_{i}}{A_{i-1}}= \prod_{i=1}^{N}  A_{\Delta}(m_{i}),
\end{equation}

where $A_{i}$ is the total amplification of a fluid quantity, $A$, of interest after a number, $i$, shocks, $A_{\Delta}$ is the individual amplification of $A$ across the $i^{\textrm{th}}$ shock, $N$ is the total number of shocks, and $\rho_{\Delta}(m_{i}) = m_{i}$ by definition.

Generally, however, each shock will introduce some turbulence anisotropy, and the amplification will depend on the turbulent velocity distributions entering each shock. Therefore, in the NIC limit, where turbulence stays anisotropic because of insufficient time to evolve between shocks, the overall amplification cannot be treated as simply. Instead, one can take the results of Ribner's LIA to calculate how the distribution of Kovasnay decomposed shear-wave modes is modified by a given shock, which, assuming no nonlinear evolution, is used as the initial condition for the next shock.

\begin{figure}
    \centering
    \includegraphics[width=0.95\columnwidth]{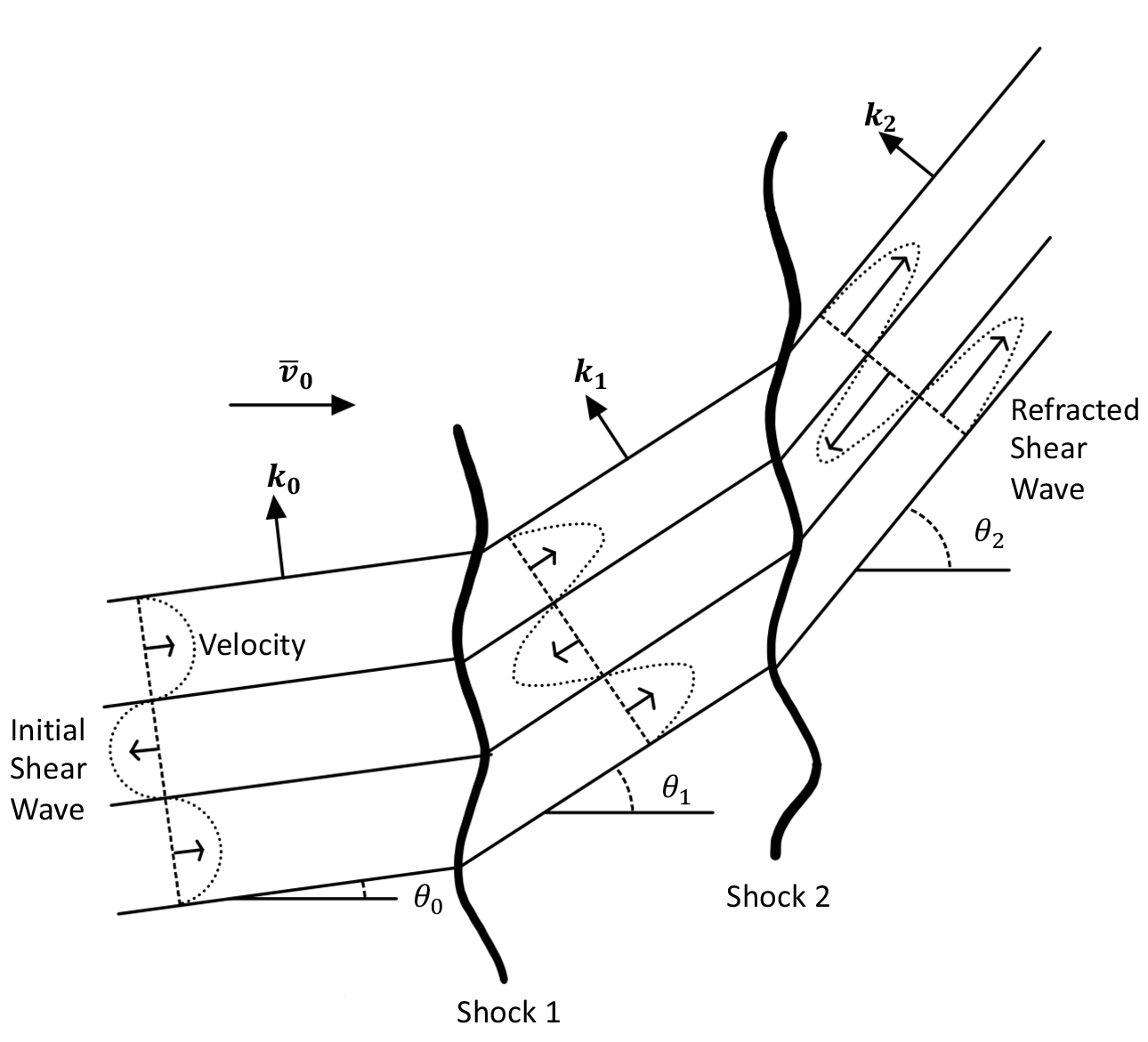}
    \caption{Refraction and amplification of a vortical shear wave as it passes through two planar shocks on the right. Based on Fig.~3 from \cite{ribner1987}, but with entropy and acoustic waves neglected here. The inclination, $\theta$, is defined here as the angle between the velocity perturbations and shock normal (or alternatively, between the wave vector, $\mathbf{k}$, and plane of the shock). This inclination, denoted as $\theta_{i}$, increases for the refracted wave across each shock.}
    \label{fig:refraction}
\end{figure}

For single-shock LIA, the Kovasnay-decomposed shear-wave modes of an initially isotropic spectrum of small amplitude, incompressible turbulence are assumed to not interact with each other across the shock. This is satisfied if $\tau_\epsilon$ is longer than the propagation time through the shock.
Linearized inviscid fluid equations and linearized Rankine-Hugoniot relations are used to solve for and relate the perturbations downstream to the upstream conditions for each initial shear wave. Thus the problem can be reduced to the interaction of each single vortical shear wave with the shock, and summed to calculate the overall turbulent amplification. It has been shown that the single-shear wave interactions with a shock can be solved in a 2D plane containing the shear wave and oblique shock \cite{ribner1987,wouchuk2009}, as in Fig. \ref{fig:refraction} (where we have added an additional shock, to illustrate the extension of the 2D treatment to the multi-shock case).

The shock has two effects on each shear wave: one is a refraction of the shear wave by shock compression, and the other is an amplification due to both compression and the perturbed, rippled shock front. In Ribner's single-shock calculation, the refracted inclination of a single shear wave is given by 
\begin{equation} \label{eq:snell}
	\tan \theta_{1} = m_{1} \tan \theta_{0},
\end{equation}

and the energy amplification, $S^2$, is calculated as a function of the shock strength and initial shear wave inclination in \cite{ribner1953}. The amplitudes of the downstream longitudinal and lateral components of the turbulent spectrum are calculated in LIA by summing the spectral densities of the refracted, amplified shear waves,

\begin{equation} \label{eq:vx}
	\overline{\tilde{v}^{2}_{x1}  } = \int | S |^{2} \frac{\cos^{2} \theta_{1}}{\cos^{2} \theta_{0}} [\tilde{v}_{x0} \tilde{v}_{x0}] \textrm{d}\mathbf{k_{0}} = \int [\tilde{v}_{x1} \tilde{v}_{x1}] \textrm{d}\mathbf{k_{1}} ,
\end{equation}

\begin{equation} \label{eq:vy}
	\overline{\tilde{v}^{2}_{y1}  } + \overline{\tilde{v}^{2}_{z1}  } = \int  \frac{| S |^{2} \sin^{2} \theta_{1} - \sin^{2} \theta_{0}}{\cos^{2} \theta_{0}} [\tilde{v}_{x0} \tilde{v}_{x0}] \textrm{d}\mathbf{k_{0}} + \overline{\tilde{v}^{2}_{y0}  } + \overline{\tilde{v}^{2}_{z0}  },
\end{equation}

where $\tilde{v}_{0}$ is the turbulent velocity upstream of the shock, $\tilde{v}_{1}$ is the downstream turbulent velocity, and $[\tilde{v}_{x} \tilde{v}_{x}]$ is the 3D spectral density of $\tilde{v}^{2}_{x} $ in wavenumber space.

Since the shear wave is only refracted in inclination across the shock and it remains in the same plane, we may assume, using axisymmetry and if there is no evolution, that the 2D treatment still applies for a second shock parallel to the initial shock. Then another refraction and amplification of the shear wave will occur across the second shock, with $S$ dependent on the strength of the second shock and the refracted inclination of the incoming shear wave. This can then be further extrapolated for N shocks, recursively using the results of Ribner's LIA to calculate the amplitude and inclination of shear waves after each shock, and therefore how the longitudinal spectral density changes between shocks,

\begin{equation} \label{eq:spec}
	\begin{split}
	[\tilde{v}_{xi} \tilde{v}_{xi}]
	&=  | S_{i} |^{2}\frac{\cos^{2} \theta_{i}}{\cos^{2} \theta_{i-1}} [\tilde{v}_{xi-1} \tilde{v}_{xi-1}] \\
	&= \frac{\cos^{2} \theta_{i}}{\cos^{2} \theta_{0}} [\tilde{v}_{x0} \tilde{v}_{x0}] \prod_{j=1}^{i}| S_{j} |^{2},
	\end{split}
\end{equation}

where $[\tilde{v}_{xi} \tilde{v}_{xi}] $ is the longitudinal spectral density of turbulence downstream of the $i^{\textrm{th}}$ shock, $\tan \theta_{i} = m_{i} \tan \theta_{i-1} = \tan \theta_{0}  \prod_{j=1}^{i} m_{j} $ is the shear wave inclination after $i$ shocks, and the shear wave amplification factor across the $i^{\textrm{th}}$ shock is $S_{i} = S(m_{i},\theta_{i-1})$.

The overall multi-shock amplification of turbulence can then be calculated from this anisotropic limiting case by summing over the final turbulence spectral densities. For initially isotropic turbulence, $\overline{\tilde{v}^{2}_{x0}  }  =\overline{\tilde{v}^{2}_{y0}  } =\overline{\tilde{v}^{2}_{z0}  } $, the $\theta_{0}$ dependency of the initial spectral density is $[\tilde{v}_{x0} \tilde{v}_{x0}](\theta_{0}) = \cos^{2} \theta_{0}$. Using this with the recursive relation, Eq.~\eqref{eq:spec}, one can find the final amplification, after N shocks, of longitudinal and lateral components of initially isotropic turbulence as

\begin{equation} \label{eq:uamp}
	\frac{\overline{\tilde{v}^{2}_{xN}  } }{\overline{\tilde{v}^{2}_{x0}  } } = \frac{3}{2} \int_{0}^{\pi /2} \cos^{2} \theta_{N} \cos \theta_{0} \prod_{i=1}^{N}| S_{i} |^{2}  \textrm{d}\theta_{0} ,
\end{equation}

\begin{multline} \label{eq:vamp}
	\frac{\overline{\tilde{v}^{2}_{yN}  } }{\overline{\tilde{v}^{2}_{y0}  } } =\frac{\overline{\tilde{v}^{2}_{zN}  } }{\overline{\tilde{v}^{2}_{z0}  } }  = 1+ \frac{3}{4} \times \\ \int_{0}^{\pi /2}  \sum_{i=1}^{N}\frac{| S_{i} |^{2} \sin^{2} \theta_{i} - \sin^{2} \theta_{i-1}}{\cos^{2} \theta_{i-1}}  [\tilde{v}_{xi-1} \tilde{v}_{xi-1}](\theta_{0}) \cos \theta_{0} \textrm{d} \theta_{0}.
\end{multline}

More generally, for initially axisymmetric turbulence, where $\alpha$ is the ratio of the energy in longitudinal modes to the energy in lateral modes such that $\overline{\tilde{v}^{2}_{x0}  }  = 2\alpha\overline{\tilde{v}^{2}_{y0}  } =2\alpha\overline{\tilde{v}^{2}_{z0}  } $, the final amplifications of each component after $N$ shocks is given by

\begin{equation} \label{eq:uampaxi}
	\frac{\overline{\tilde{v}^{2}_{xN}  } }{\overline{\tilde{v}^{2}_{x0}  } } =  \frac{\int_{0}^{\pi /2} 
 \frac{\cos^{2} \theta_{N}}{\cos \theta_{0}} [\tilde{v}_{x0} \tilde{v}_{x0}]  \prod_{i=1}^{N}| S_{i} |^{2}  \textrm{d}\theta_{0}}{\int_{0}^{\pi /2} 
 \cos \theta_{0}  [\tilde{v}_{x0} \tilde{v}_{x0}]  \textrm{d}\theta_{0}} ,
\end{equation}

\begin{multline} \label{eq:vampaxi}
	\frac{\overline{\tilde{v}^{2}_{yN}  } }{\overline{\tilde{v}^{2}_{y0}  } } =\frac{\overline{\tilde{v}^{2}_{zN}  } }{\overline{\tilde{v}^{2}_{z0}  } }  = 1+ \\ \frac{\alpha \int_{0}^{\pi /2}  \sum_{i=1}^{N}\frac{| S_{i} |^{2} \sin^{2} \theta_{i} - \sin^{2} \theta_{i-1}}{\cos^{2} \theta_{i-1}}  [\tilde{v}_{xi-1} \tilde{v}_{xi-1}](\theta_{0}) \cos \theta_{0} \textrm{d} \theta_{0}}{\int_{0}^{\pi /2} 
 \cos \theta_{0}  [\tilde{v}_{x0} \tilde{v}_{x0}]  \textrm{d}\theta_{0}}.
\end{multline}

We use the amplification calculated by this model for initially isotropic turbulence in both IC and NIC regimes to explore general effects of shock strength and the feedback of generated anisotropy on the multi-shock compression of turbulence in the next section.

\section{Results}
\label{sec:results}

\subsection{Trends with shock strength}
\label{sec:trends}

In some practical situations, such as compression of ICF capsules, there may be a desired final density from shock compression, and the amplitude of any generated turbulence in the final state may be of interest. To generically understand the influence of shock strength on TKE amplification in the context of multiple shocks in the NIC or IC limits, we first consider a series of $N$ shocks of equal strength. Suppose the final density amplification, $\rho_{N} $, is kept fixed, such that the number, $N$, of identical shocks is determined by the choice of shock strength. The final density amplification is given by Eq.~\eqref{eq:gen_amp} for $A=\rho$ and $m_{\rm i}=m$ for all shocks, constraining $N$ as 

\begin{equation} \label{eq:N}
	N =  \frac{\log \rho_{N}}{\log m}.
\end{equation}

Using the definition in Eq.~\eqref{eq:e_def} and conservation of mass, the final TKE amplification after N shocks is given by 
\begin{equation} \label{eq:energy_amp_defn}
	\frac{E_{N}}{E_{0}}= \frac{\rho_N}{\rho_0} \frac{V_N}{V_0} \frac{\overline{\mathbf{\tilde{v}}_N^{2}}}{\overline{\mathbf{\tilde{v}}_0^{2}}}=\frac{\overline{\mathbf{\tilde{v}}_N^{2}}}{\overline{\mathbf{\tilde{v}}_0^{2}}}.
\end{equation}

First consider the IC limit, where turbulence evolves and returns to isotropy between shocks. Then the final TKE amplification can be found from Eq.~\eqref{eq:gen_amp}, for $A=E$, as

\begin{equation} \label{eq:iso_energy}
	\frac{E_{N}}{E_{0}}= E_{\Delta}^N,
\end{equation}

where the amplification across any given shock $E_{\Delta}(m)=E_i/E_{i-1}=\overline{\mathbf{\tilde{v}}_i^{2}}/\overline{\mathbf{\tilde{v}}_{i-1}^{2}}$ is identical between shocks and can be computed from single-shock LIA via Eqs.~\eqref{eq:vx} and~\eqref{eq:vy}.

Substituting for $N$ in Eq.~\eqref{eq:iso_energy} using Eq.~\eqref{eq:N}, $E_{N}$ can be expressed as

\begin{equation} \label{eq:eff_form}
	\frac{E_{N}}{E_0} = \left( E_{\Delta}^{1/\log m    } \right)^{\log \rho_{N}}.
\end{equation}

In the form of Eq.~\eqref{eq:eff_form}, it is apparent that general trends of how the overall TKE amplification ($E_{N}$) depend on single-shock density jumps ($m$) can be drawn from the bracketed quantity, and the effect of increasing the final density is to magnify any such trends through repeated shocks.

\begin{figure}
    \centering
    \includegraphics[width=0.95\columnwidth]{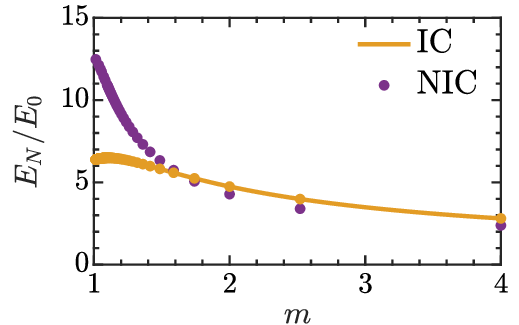}
    \caption{Final TKE amplification after $N$ shocks of equal strength, $m$, to reach a final density amplification of $\rho_{N}/\rho_{0}=16$. The IC limit is shown by the orange points representing integer numbers of shocks, filled in between by a orange line for fractional values of $N$ (physically, the number of shocks must be an integer). The NIC limit is given by the purple points.}
    \label{fig:shock_eff}
\end{figure}

The final TKE amplification for the IC limit as a function of shock strength is given by the orange line in Fig.~\ref{fig:shock_eff}. It reaches a maximum at a shock strength of $m~\approx 1.1$, and approaches a minimum towards the strong shock limit. The exact position of this maximum can be determined from Eq.~\eqref{eq:eff_form} as

\begin{equation} \label{eq:max}
	\frac{\textrm{d} E_{N}/E_{0}}{\textrm{d} m }  =    E_{N} \ln \rho_{N}    \left( \frac{\textrm{d} E_{\Delta}  / \textrm{d} m  }{E_{\Delta} \ln m} - \frac{\ln E_{\Delta}}{m (\ln m)^{2}}     \right) = 0,
\end{equation}

by equating the bracketed quantity to zero. This shock strength for maximally efficient amplification does not depend on the final density. Rather, it depends only on what is used for the single-shock TKE jump $E_{\Delta}$, for which we used Ribner's LIA here (Eqs.~\eqref{eq:vx} and~\eqref{eq:vy}), which itself only depends on $\gamma$ and $m$. Thus for any fixed final density, choosing shocks of strength $m \approx 1.1$ will maximally amplify turbulence for $\gamma=5/3$ in the IC limit.

Next consider the NIC limit, where turbulence does not return to isotropy. For sufficiently weak turbulence, the jumps in mean background quantities will be unchanged from the IC limit, described by substituting the single-jump Rankine-Hugoniot relations into Eq.~\eqref{eq:gen_amp}. However the TKE amplification after $N$ shocks departs from Eq.~\eqref{eq:gen_amp}, and is instead given by Eq.~\eqref{eq:uamp} and Eq.~\eqref{eq:vamp}. Therefore, $E_{N}$ depends on $m$ through the history of how each decomposed shear wave is amplified at each shock, $S_{i}(m,\theta_{i})$, and refracted, $\tan \theta_{i} = m^{i} \tan \theta_{0}$. Because this differential amplification results in increasingly anisotropic turbulence with each successive shock, the overall amplification $E_{N}$ also varies non-trivially with the total number of shocks $N$ and final density $\rho_{N}$. Regardless, an example for a fixed final density amplification of factor $16$ is plotted in purple in Fig. \ref{fig:shock_eff}. In this NIC limit, the TKE amplification monotonically decreases with increasing shock strength, similar to the isotropic case but without extrema in the trend. Other choices of final density are not shown, however a similar monotonic decrease is observed. For a given final background state in both IC and NIC limits, launching a series of weaker shocks generally results in greater amplification of turbulence than utilizing strong shocks, with the exception of the maximum at $m~\approx 1.1$ for the IC limit.

\begin{figure}
    \centering
    \includegraphics[width=0.95\columnwidth]{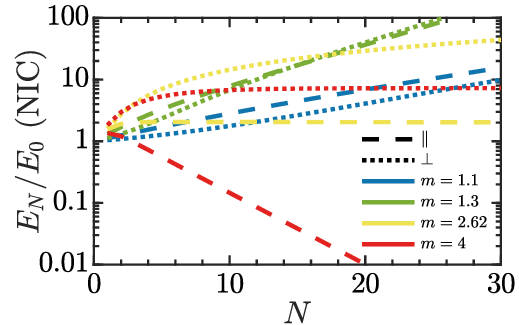}
    \caption{Total amplification of longitudinal (parallel to shock propagation direction, dashed) and lateral (perpendicular to shock propagation, dotted) components of TKE, in the NIC limit, after launching multiple shocks of strengths a) $1.1$ (blue) b) $1.3$ (green) c) $2.62$ (yellow) c) $4$ (red). Note that different final densities are reached by each strength, for the same number of shocks.}
    \label{fig:Nshock}
\end{figure}

As the turbulence will generally be anisotropic in the NIC limit, it may be instructive to examine the general trends with shock strength for amplification of individual components. This is done in Fig.~\ref{fig:Nshock} for launching increasing numbers of identical shocks of strength $m=1.1$, $m=1.3$, $m=2.62$, or $m=4.0$. $m=1.3$ is chosen as a relatively weak shock, and because it maintains relative isotropic amplification of both components in the single-shock case. For the series of $m=1.3$ shocks, the overall amplification stays closer toward isotropy, and monotonically increases with more shocks. The $m=1.1$ case similarly shows a monotonic increase in TKE amplification with shock number, albeit preferentially longitudinal and reduced relative to the $m=1.3$ case for an equivalent number of shocks due to the weaker shock strengths. As the strengths of each shock are increased, the stronger refraction results in preferential lateral amplification over the longitudinal component. After enough shocks, this can result in the amplification of the longitudinal component stagnating, and even being suppressed by additional shocks. This can occur for fewer than $30$ shocks if the strength of each shock is greater than $m=2.61$. The longitudinal amplification for a series of $m=2.62$ shocks is shown by the yellow dashed line in Fig.~\ref{fig:Nshock}, which reaches a maximum at $6$ shocks, and thereafter the longitudinal amplitude decreases. A more drastic case is shown in red for the strong shock limit, $m=4$, where the longitudinal component is suppressed strongly by all shocks following the first. In these cases where longitudinal amplification is limited, the ability of shocks to amplify the lateral component reaches a plateau.

\subsection{Polytropic index/Quasi-EOS}
\label{sec:eos}

A more general understanding for how turbulence behaves and partitions energy under compression from multiple shocks can be acquired by constructing a quasi-equation of state, similarly to \cite{davidovits2019}, relating TKE to a pressure. From dimensional arguments, this is

\begin{equation} \label{eq:turbp}
	p_{turb} \propto E_{TKE} = \frac{1}{2} \rho \overline{\tilde{v}^{2} }.
\end{equation}

This expression for the pressure can then be used to find a polytropic index for the compression of turbulence under a series of shocks:

\begin{equation} \label{eq:index}
	n(N) = \frac{\partial \log p_{N} }{\partial \log \rho_{N}},
\end{equation}

where $n(N)$ is the polytropic index after $N$ shocks, and $p$ is a pressure that could be thermal or from TKE. Generally, $n(N)$ may depend on the history of compression (the order, number and strength of all shocks up to shock $N$).

Consider again launching identical shocks of equal strength. In the IC limit, with isotropization between each shock,  we can use Eq.~\eqref{eq:gen_amp} for $N$ shocks of strength $m$, together with Eq.~\eqref{eq:index} to find the index,

\begin{equation} \label{eq:iso_ind}
	n(N) = \frac{\partial \log p_{\Delta}^{N} }{\partial \log m^{N}} =   \frac{\partial \log p_{\Delta} }{\partial \log m},
\end{equation}

which is only a function of the single-shock jumps for density and turbulent pressure. Therefore, by taking a weak-shock expansion of single-shock LIA, one can also find a weak shock limit for the index. We find, to first order in $(m-1)$, the amplification of the vortical turbulence components in this limit to be

\begin{equation} \label{eq:linearu}
	\frac{\overline{\tilde{v}^{2}_{xi}  } }{\overline{\tilde{v}^{2}_{xi-1}  } }  = 1 + \frac{6}{5} (m-1) + O[(m-1)^{2}],
\end{equation}

\begin{equation} \label{eq:linearv}
	\frac{\overline{\tilde{v}^{2}_{yi}  } }{\overline{\tilde{v}^{2}_{yi-1}  } }  = \frac{\overline{\tilde{v}^{2}_{zi}  } }{\overline{\tilde{v}^{2}_{zi-1}  } } = 1 + \frac{2}{5} (m-1) + O[(m-1)^{2}].
\end{equation}

Combining with Eq.~\eqref{eq:turbp} to obtain the vortical turbulent pressure jump, Eq.~\eqref{eq:iso_ind} gives a weak shock index of $n=5/3$. A similar weak shock expansion of the thermal pressure also yields the same polytropic index of $n=5/3$, corresponding to the adiabatic limit.

\begin{figure}
    \centering
    \includegraphics[width=0.95\columnwidth]{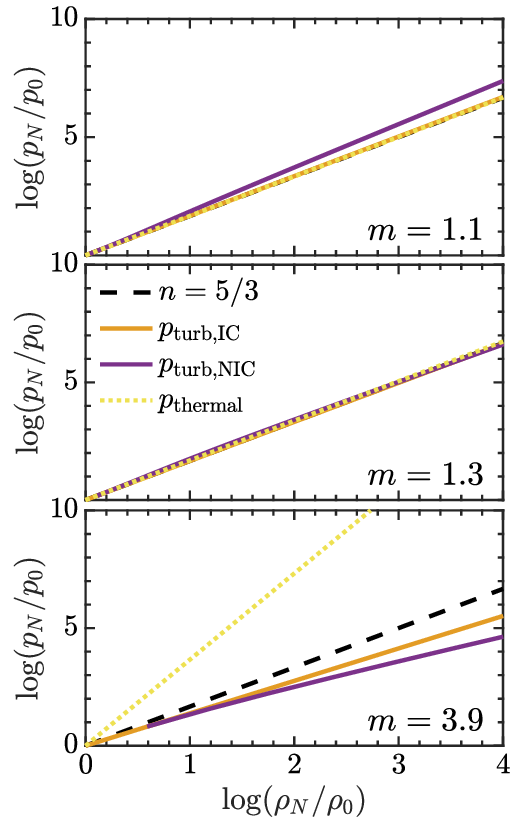}
    \caption{Total thermal (yellow, dotted) and turbulent (orange for IC, purple for NIC) pressure jumps on a log-log (base $10$) scale versus total density jump for multiple equivalent shocks of strengths $m=1.1$ (top), $m=1.3$ (middle), and $m=3.9$ (bottom). Adiabatic prediction of thermal pressure ($n=5/3$) shown by black dashed line.}
    \label{fig:quasiWeak}
\end{figure}

As has been discussed in the previous section, due to the feedback of anisotropy, the amplification of TKE across any given shock in the NIC limit depends non-trivially on previous shocks. Therefore the NIC polytropic index is also a function of the the history of shocks. The instantaneous value of $n$ for a given shock in a series can be inferred from the TKE amplification by taking the gradient of the curves in Fig.~\ref{fig:quasiWeak}, which shows $\log (p_N/p_0)$ plotted against $\log (\rho_N/\rho_0)$ for increasing numbers of shocks, $N$, of strengths $m=1.1$, $m=1.3$, $m=3.9$ in both the NIC and IC limits.

Inspection of the gradients for the IC limit with $m=1.1$ and $m=1.3$ in Fig.~\ref{fig:quasiWeak} reveals that the weak-shock IC limit tends towards an index of $n=5/3$, as predicted from the weak-shock expansion of LIA. Therefore, in the limit of shocks being sufficiently weak and isotropization between shocks (IC), a compression via repeated shocks corresponds to a 3D, isotropic, adiabatic metric compression. Curiously, the NIC limit with m=1.3 also shows an index close to $n=5/3$, owing to the relative isotropy of amplification predicted by LIA in this case, even with repeated shocks, as seen in Fig.~\ref{fig:Nshock}. However, conversely, the weak-shock NIC limit exhibits a steeper slope, with a super-adiabatic index $n\approx1.83>5/3$, requiring more energy to compress.

Increasing the shock strength results in a greater proportion of heating than compared to compression and turbulence amplification. This is perhaps expected from single-shock jump trends, where the density and TKE jumps are bounded, whereas the temperature jump is unbounded in the strong shock limit. We see this manifesting in the IC and NIC effective indices in Fig.~\ref{fig:quasiWeak}, where the index for the turbulent pressure decreases with shock strength, while that of the temperature can increase unbounded in the strong shock limit. Thus, in the strong shock limit nearly all the compression energy does work against the thermal pressure and goes into heating. The effects of increasing shock strength are more adverse in the NIC limit, where the TKE amplification can become highly anisotropic when strong shocks are launched repeatedly, as in Fig.~\ref{fig:Nshock}.
The effect of this increasing anisotropy on the instantaneous index can be inferred from the purple line in Fig.~\ref{fig:quasiWeak}, whose gradient decreases with an increasing number of shocks in the strong-shock case. As in Fig.~\ref{fig:shock_eff}, a series of stronger shocks does not amplify TKE as efficiently, and so the strong-shock index decreases as the turbulence becomes increasingly anisotropic with each shock, such that minimal work is done against TKE by the compression.

\subsection{Shock ordering}
\label{sec:ordering}

In some compression scenarios, such as achieving a low adiabat in ICF, multiple shocks of different strengths may be used and the order selected so as to minimize the required energy or to ensure stability \cite{lindl1995}.
Imagine such a scenario with a desired compression level after multiple shocks (as previously considered in Section~\ref{sec:trends}) but now with shocks of possibly unequal strength. Because the final compression is fixed, the mean quantities again do not depend on the shock order.
On the other hand, given the previous discussion on the dependence of TKE amplification in the NIC limit on the history of shocks, the amplification of turbulence may be sensitive to a change in the order in which the shocks are launched.

\begin{figure}
    \centering
    \includegraphics[width=0.8\columnwidth]{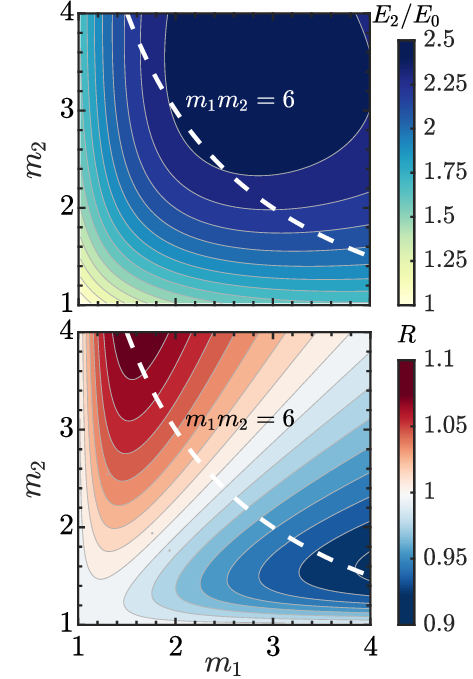}
    \caption{Top: Contour plot of TKE amplification,  after two shocks, $E_2/E_0$, for an initially isotropic distribution of turbulence, in the NIC limit. Strength of initial shock is given on the horizontal axis, and of the second shock on the vertical axis. Bottom: Corresponding contour plot of ordering ratio, $R$. The maximal difference is a factor of $1.084$ for an initial weaker shock of $m=1.58$, followed by a strong shock of $m=4$. An example contour of constant final density amplification of factor $6$ is shown by the white dashed line in both plots.}
    \label{fig:total_order}
\end{figure}

The top panel of Fig.~\ref{fig:total_order} shows a contour plot of the the amplification of TKE after a shock of strength $m_1$ (horizontal axis), followed by a shock of strength $m_2$ (vertical axis), for initially isotropic turbulence in the NIC limit.
The amplification is asymmetric between the two shocks. This can be seen from the shapes of the contours or following a path of constant final density (e.g. $m_1 m_2 =6$), along which $E_2/E_0$ is seen to be greater for the half of the path at $m_1<m_2$. Furthermore, TKE amplification reaches a maximal value of $E_2/E_0=2.5$ for shocks of strengths $m_1=2.7$ and $m_2=3.4$, but reversing the order of shocks such that $m_1=3.4$ and $m_2=2.7$ reduces the amplification to $E_2/E_0=2.43$.

This asymmetry in TKE amplification with respect to shock order is more apparent from calculating an ordering ratio, $R$, of the amplification with one ordering of shocks versus the other,

\begin{equation} \label{eq:ratio}
	R(m_{1},m_{2}) = \frac{E(m_{1},m_{2})}{E(m_{2},m_{1})},
\end{equation}

where $E(m_{1},m_{2})$ is the amplification of TKE after a shock of strength $m_{1}$, followed by $m_{2}$. Values of $R>1$ mean the original ordering of shocks leads to greater amplification than if the order was reversed.

The ordering ratio, $R$, corresponding to the TKE amplification example in the top panel of Fig.~\ref{fig:total_order}, is similarly plotted against $m_1$ and $m_2$ in the bottom panel of Fig.~\ref{fig:total_order}. As expected, $R$ is ``anti-symmetric" around the diagonal, $m_1 = m_2$, in the sense that $R(m_1,m_2) = 1/R(m_2,m_1)$. That the plot is red in the upper diagonal and blue in the lower diagonal shows that $R>1$ for $m_1<m_2$, and $R<1$ for $m_1>m_2$. Therefore, for two given shock strengths, the TKE amplification of initially isotropic turbulence by the two shocks is maximized if the weaker shock is launched first.

The maximal difference in amplification due to shock ordering is a factor of $R_{max}=1.084$, if the order of two shocks of strength $m=1.58$ and $m=4$ are swapped. While $R_{max}=1.084$ may be small for the isotropic distribution of turbulence, turbulence in experiments and nature can often be highly anisotropic and the difference in amplification due to shock order can potentially be enhanced. To understand this and as a general approach, it is important to examine how the sensitivity to shock order for individual shear waves can be much greater for certain initial inclinations.

\begin{figure*}
    \centering
    \includegraphics[width=0.956\textwidth]{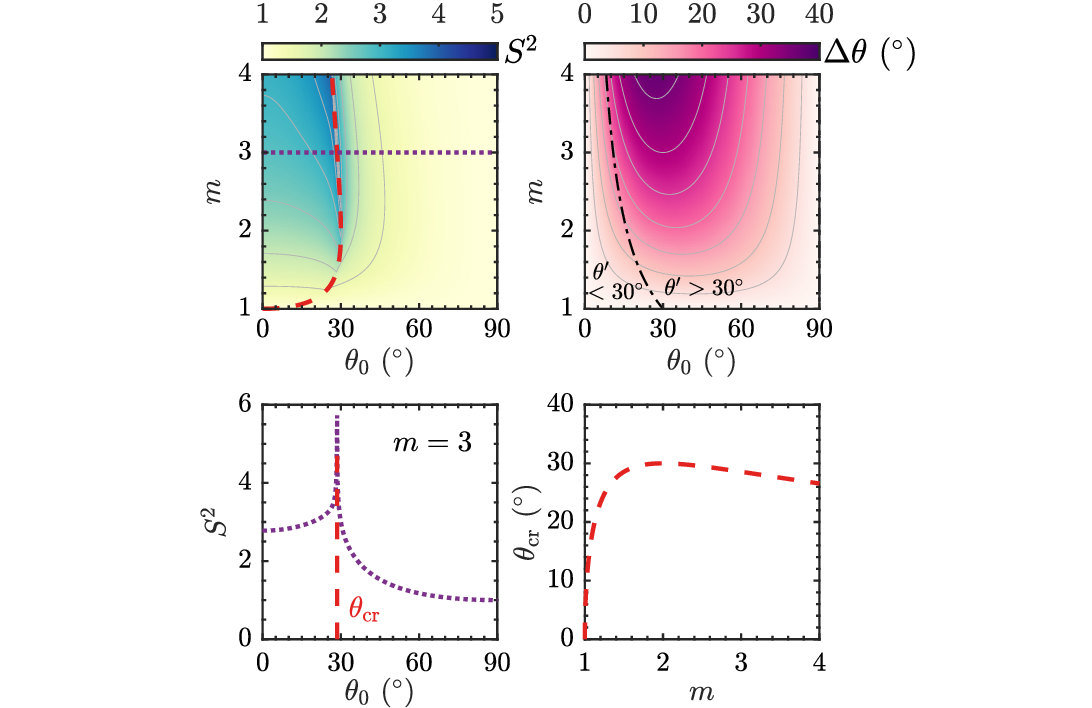}
    \caption{Top: $2$D plots of amplification of energy ($S^2$, top left) and refraction (change in inclination, $\Delta \theta=\theta'-\theta_0$, top right) of a single shear wave of initial inclination $\theta_0$, by a single shock of strength $m$. Isocontours of the respective quantities are also plotted. The dash-dotted black line (top right) is the isocontour for $\theta'=\theta_{\rm cr}=30^\circ$. To the right of this isocontour, the refracted wave will exceed the critical inclination for any additional shocks.
    Bottom left: Amplification of shear wave energy by a single shock of strength, $m=3$ (path along purple dotted line in top plot), plotted against initial inclination of the shear wave. A vertical red dashed line is plotted at the the critical inclination for $m=3$, $\theta_{\rm cr}=28.56^{\circ}$.
    Bottom right: Critical initial inclination, $\theta_{\rm cr}$, of a shear wave for a shock of strength $m$. This is also plotted by the red dashed line in the top left panel.
    }
    \label{fig:crit_angle}
\end{figure*}

The LIA dependence of the single-shock energy amplification of an individual shear wave on its inclination is shown in the left panels of Fig.~\ref{fig:crit_angle}, which plot $S^2$ against initial shear wave inclination $\theta_0$ and shock strength $m$ (top left), and versus $\theta_0$ for a fixed $m=3$ (bottom left). There exists a sharp peak in amplification at a ``critical inclination",
\begin{equation} \label{eq:thcr}
	\theta_{\rm cr} = \arctan\sqrt{\frac{(\gamma+1)(m-1)}{2m^2}},
\end{equation}

for which the resultant post-shock flow in the steady-flow frame (transformation velocity increases with inclination) reaches the sound speed \cite{ribner1954}. At $\theta_{\rm cr}$, the amplification in shear wave energy can be up to twice as large than at the smallest inclinations. This peak can be seen in the $m=3$ example, where $\theta_{\rm cr}=28.56^\circ$. With increasingly transverse inclinations beyond the critical inclination, the amplification is rapidly suppressed. While the critical inclination of a shear wave depends on the strength of the shock, as plotted for Eq.~\eqref{eq:thcr} in the bottom right panel of Fig.~\ref{fig:crit_angle}, it is generally a preferentially longitudinal inclination that reaches a maximum $\theta_{\rm cr,max}=30^{\circ}$ for a shock of strength $m=2$ and $\gamma=5/3$. For a large range of shock strengths above $m=1.126$, $\theta_{\rm cr}$ lies between $20^{\circ}< \theta_{\rm cr} < 30^{\circ}$.

The corresponding refraction of a shear wave of inclination $\theta_0$ through a single shock of strength $m$ is given by the change in its inclination, $\Delta \theta$, plotted in the top right panel of Fig.~\ref{fig:crit_angle}. This refraction is greater for stronger shocks, predominantly for intermediate inclinations around typical $\theta_{\rm cr}$. The black dash-dotted line separates the plot into a region where the post-shock inclination, $\theta'=\Delta \theta + \theta_0$, is below the maximum critical inclination, $\theta'<\theta_{\rm cr,max}$ (left of the line), or exceeds the maximum $\theta'>\theta_{\rm cr,max}$ (right of the line). This left region demarcates the parameter space of initial shear-wave inclination and initial shock strength for which the refracted shear wave could still experience critical effects for subsequent shocks.

\begin{figure*}
    \centering
    \includegraphics[width=0.956\textwidth]{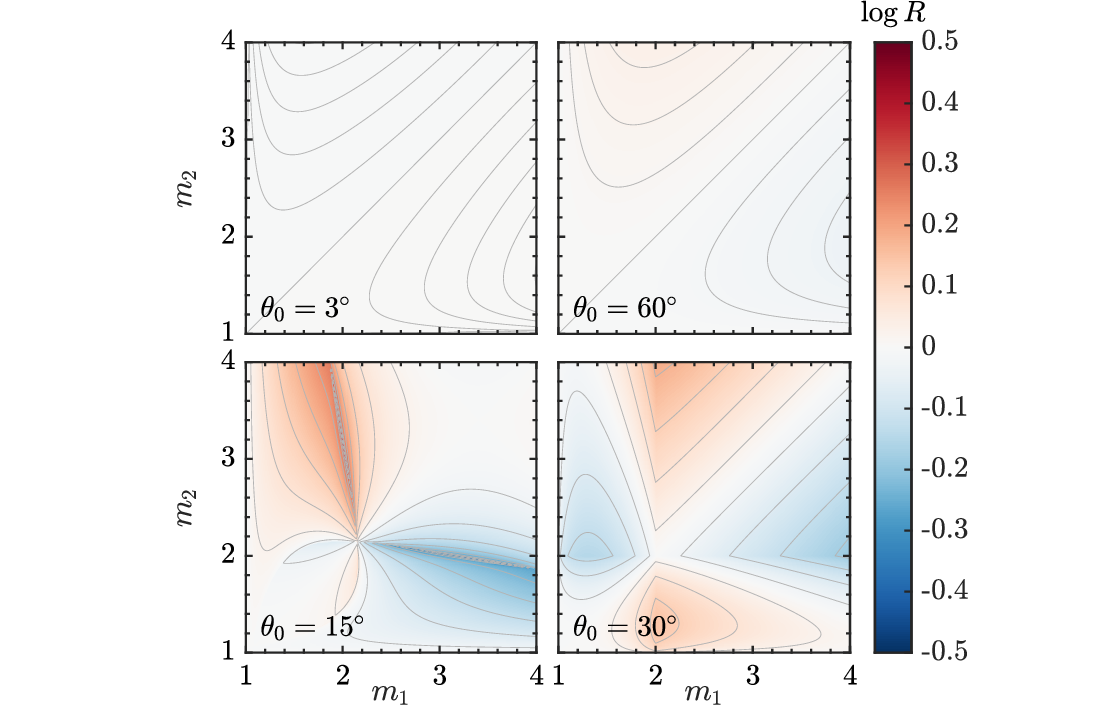}
    \caption{$2$D logarithmic (base $10$) plots of ordering ratio $R$, for the two-shock amplification of shear waves of varying initial inclination between each subplot ($3^\circ$, $60^\circ$, $15^\circ$, and $30^\circ$), in the NIC limit. Strength of initial shock is given on the x axis, and of the second shock on the y axis. Isocontours of $\log R$ are overlaid.}
    \label{fig:shear_order}
\end{figure*}

We now consider such a two-shock system, where the strengths of each shock are denoted by $m_{1}$ and $m_{2}$, and we can choose to launch either $m_{1}$ or $m_{2}$ first, followed by the remaining shock. We will denote the initial inclination of the shear wave to be $\theta$ and the inclination after the first shock to be $\theta'$. Then $\theta'$ will depend on which of the shocks is launched first, $\theta'(m_{i},\theta) = \arctan [m_{i} \tan(\theta) ]$. The two-shock energy amplification is given by the product of the amplifications from each shock, which depend on the incoming shear wave inclinations and individual shock strengths, $S^{2}_{12} = S^{2}(m_{1},\theta) S^{2}(m_{2},\theta'(m_{1},\theta))$. If the order of the shocks is reversed, then $S^{2}_{21} = S^{2}(m_{2},\theta) S^{2}(m_{1},\theta'(m_{2},\theta))$. The ordering ratios for a single shear wave, $R(m_{1},m_{2},\theta)= S^{2}_{12}(\theta) / S^{2}_{21}(\theta)$, therefore expresses the difference in amplification of shear-wave energy due to shock ordering.

If the initial shear wave inclination is too longitudinal with respect to the shock, then the shear wave cannot be refracted enough to reach the range of critical inclinations shown in the bottom right panel of Fig.~\ref{fig:crit_angle} for the second shock. For example, shear waves with $\theta_0 \approx 0$ (top right panel) are not refracted to the $20^{\circ}< \theta' < 30^{\circ}$ range. Likewise, if the initial inclination surpasses $30^{\circ}$ ($\theta>\theta_{\rm cr,max}$), then it is beyond the critical inclination for both of the shocks.

Ordering ratios for inclinations far outside the range of criticality, such as $3^{\circ}\ll \theta_{\rm cr}$ and $60^{\circ}\gg \theta_{\rm cr}$, are plotted against $m_1$ and $m_2$ in the top panels of Fig.~\ref{fig:shear_order}. In the far-subcritical regime ($\theta \ll 20^{\circ}$ e.g. $3^{\circ}$), the amplification is maximized with a stronger initial shock that refracts the shear wave more strongly to (still subcritical) inclinations where the subsequent shock amplification is slightly increased. For example, at $\theta=3^{\circ}$, amplification is slightly enhanced by up to a maximum of $0.58\%$ for $m_1=4$ and $m_2=1.6$. In the supercritical regime ($\theta \gg 30^{\circ}$ e.g. $60^{\circ}$), the amplification is instead maximized by launching the weaker shock first to minimize refraction such that the inclination of the shear wave does not depart as far from $\theta_{\rm cr}$ for the subsequent shock. For $\theta=60^{\circ}$, amplification can be up to a maximum of $8.26\%$ greater for $m_1=1.93$ and $m_2=4$. In both of these non-critical regimes, because the amplification is not greatly enhanced by critical effects for either shock, the differences in TKE amplification due to shock ordering are minimal.

Conversely, closer to the range of critical angles ($ \theta \lesssim 30^{\circ}$), differences in amplification due to ordering can be substantially larger. Amplification of shear wave energy at example inclinations of $15^{\circ}<\theta_{\rm cr}$ and $30^{\circ} = \theta_{\rm cr,max}$ is plotted in the bottom panels of Fig.~\ref{fig:shear_order}. In the near-subcritical regime ($ \theta \lesssim 20^{\circ}$), a weaker first shock can refract the shear wave to inclinations that are critical for a stronger second shock, maximizing the amplification substantially compared to the reverse configuration that greatly minimizes the amplification. An example of this is shown for an initial inclination of $15^{\circ}$ in the bottom left panel of Fig.~\ref{fig:shear_order}. Most of the space for $m_1<m_2$ results in positive values of $R$, with a maximum of $R=3.19$: a factor $3.19$ greater energy amplification for a shear wave, initially inclined at $15^{\circ}$, can be achieved by launching a weaker shock of strength $m_1=  1.94$ followed by a shock of strength $m_2= 3.54$ than if the shocks were reversed in order.

For inclinations close to the maximum critical inclination $\theta \approx 30^{\circ}$, refraction by even relatively weak shocks (see Fig.~\ref{fig:crit_angle}, top right) will result in a supercritical inclination for a following stronger shock and suppress the overall amplification. Therefore, in this regime, the amplification can instead be maximised by choosing the strength of the first shock such that the initial inclination is critical, regardless of the strength of the second shock. Reversing the order of the shocks, such that the first non-critical shock refracts the initial shear wave beyond the critical inclination of the other shock, will not produce as large an amplification. An example of this regime is shown for $\theta =30^{\circ}$ in the bottom right panel of Fig.~\ref{fig:shear_order}. An inclination of $\theta =30^{\circ}$ is critical for a shock of strength $m=2$. Therefore, the amplification when launching a critical first shock of $m_1=2$ followed by a strong shock of $m_2=4$ is $54.1\%$ greater than launching a first shock of $m_1=4$ followed by $m_2=2$.

We therefore see for these inclinations where critical angle effects come into play in the multi-shock problem, the differences in shear wave amplification are substantially more extreme, with over a factor of three difference being possible for an initial inclination of $15^{\circ}$. This is in stark contrast to the few percent differences observable for inclinations far from critical angle effects, and for an initially isotropic distribution of turbulence. For the latter case, the large differences due to near-critical shear wave components of the turbulence are smoothed out by the other far-critical components.

\section{Discussion}
\label{sec:discussion}

We now discuss the relevancy of the significant findings for certain example applications. It was mentioned and explored in Section~\ref{sec:trends} (particularly in Fig.~\ref{fig:shock_eff}) that for certain scenarios, such as in ICF, the final density from shock compression may be fixed, and the amplitude of turbulence in the final state could be of interest.

There could be scenarios where a final state with maximal turbulence is desired, for example a fast ignition scheme where compression energy can be transferred to TKE and released in a final compressed state \cite{davidovits2016}. In a variation of such a fast ignition scheme, the TKE may in and of itself be advantageous to directly increase the hot spot fusion reactivity \cite{fetsch2024}.
In such a setup, our analysis has shown launching a series of weaker shocks to be optimal for TKE amplification, especially so for the NIC limit. This is generally true in the IC limit too, albeit with shocks of strength $m=1.1$ being optimal for $\gamma=5/3$. A series of weaker shocks also has the benefit of minimizing the proportion of compression energy used on heating, with both turbulent and thermal pressures exhibiting adiabatic behavior in the weak-shock limit.

On the other hand, if turbulence in the final state needs to be minimized such as to reduce mix, then a series of fewer, stronger shocks should be chosen, for the same final compression. This is particularly the case for the NIC limit, which also has the additional effect of magnifying the turbulent anisotropy. This anisotropy reduces further amplification and favors components of turbulence transverse to the shock while suppressing the longitudinal components. Therefore in this limit for an ICF capsule, the longitudinal turbulent fluctuations will be suppressed by a series of strong shocks, further reducing mixing in the radial direction and associated cooling of the hotspot. However, these turbulence considerations of course need to be balanced against the other design constraints, since stronger shocks result in a greater proportion of heating.

The most major difference from the IC limit is the effect of shock ordering in the NIC limit. We have shown how while background mean quantities are agnostic to the order of shocks for weak turbulence, the TKE amplification can be sensitive to shock order. This sensitivity to shock ordering may be an additional factor to consider when choosing shocks in compression scenarios such as ICF.

For just the two-shock case with initially isotropic turbulence, differences of up to only $8.4\%$ are possible from swapping the order of the two shocks, with the larger turbulence amplifications coming from launching the weaker shock first. Such differences could be changed with the addition of more shocks, or with different initially anisotropic distributions of turbulence that are often more physically common than the ideal case of isotropic turbulence. We have shown how this can be done for the two-shock case with individual shear waves. The sensitivity to shock ordering strongly depends on the initial inclination of the shear wave with respect to the shock, with up to a factor $3.2$ difference in TKE amplification for an inclination of $15^{\circ}$.

Alternatively, this sensitivity presents a novel diagnostic for shock ordering. One could imagine a situation where there exists an anisotropic distribution of turbulence consisting mainly of shear waves within a range of near-critical inclinations. It would then be expected that if it was passed through two shocks, one with $m_{1}\approx 2$ and $m_{2}\approx 4$, and then again with the orders reversed, one could determine the order in which they were launched based on the turbulent amplification. This would typically be difficult to determine from comparing the mean quantities in the initial and final states.

While the results presented in Section~\ref{sec:results} assume initially isotropic turbulence, the consequences of feedback of generated anisotropy in the NIC limit provides an intuition on how our other results would change for initially anisotropic turbulence. For example, from the discussion on Fig.~\ref{fig:Nshock} and the enhanced amplification of sub-critical shear waves seen in Fig.~\ref{fig:crit_angle}, turbulence that consists of dominantly longitudinally-oriented shear waves could be expected to be amplified more strongly than isotropic turbulence.

This type of axisymmetric turbulence is naturally generated in simulations of ICF implosions including perturbation sources, such as in the first NIF experiment to achieve an igniting fusion plasma \cite{abu-shawareb2022,zylstra2022,kritcher2022}.
There are three shocks in this experimental design. In 2D simulations including initial density perturbations in the ablator from its grain structure \cite{davidovits2022b}, the first shock (strength $m_0\approx2$) generates an initial vorticity field which is strongly biased towards perturbations that are longitudinal relative to the second and third shocks of strengths $m_1\approx1.75$ and $m_2\approx2$ respectively.

Taking the post-shock vorticity spectrum as a function of inclination from the simulation in \cite{davidovits2022b}, and assuming, given axisymmetry, an equal partition between energy in longitudinal and lateral modes in $3$D as was in the $2$D simulation, multi-shock amplifications of this initially axisymmetric turbulence can be calculated using Eqs.~\eqref{eq:uampaxi} and \eqref{eq:vampaxi}. For the whole shock-strength parameter space, we find the maximum TKE amplification to be $E_2/E_0=7.97$ (for $m_1=3.8$ and $m_2=4$) in the NIC limit, over three times greater than the maximum amplification of $E_2/E_0=2.50$ available to initially isotropic turbulence. We predict the amplification of TKE after shocks $m_1=1.75$ and $m_2=2$ from the experimental design to be a factor of $E_2/E_0=4.60$. For initially isotropic turbulence, we find amplification factors of $E_2/E_0=2.11$ in the NIC limit, and $E_2/E_0=2.06$ in the IC limit. Therefore, we predict the expected amplification in this design to be twice as large for this example of realistic turbulence with a longitudinal bias than for isotropic turbulence, consistent with expectations from our results. For an initial turbulent velocity of $\sim0.5$~km/s from \cite{davidovits2022b}, we expect a final turbulent velocity of $\sim0.5\sqrt{4.6}=1.07$~km/s.

The predicted ideal jumps in background density and temperature (ignoring rarefractions between shocks that occur in the actual ICF design) under these two shocks are $\rho_2/\rho_0=3.5$ and $T_2/T_0=3.68$, both less than the energy amplification of the anisotropic turbulence. From our quasi-EOS analysis in Section \ref{sec:eos}, this indicates that this anisotropic turbulent pressure would behave super-adiabatically under compression for this experiment, more so than the thermal pressure. Since the turbulence is more anisotropically biased in the direction of shock compression, more work needs to be done against the turbulent pressure.

The difference in amplification under swapping the order of the two shocks is relatively negligible for this anisotropic distribution of turbulence, only $1.10\%$, less than the difference of $1.33\%$ for isotropic turbulence. Even for unconstrained choices of shock strength, the maximal factor difference due to ordering is only $R_{\rm max}=1.062$ ($6.2\%$), less than the $R_{\rm max}=1.084$ ($8.4\%$) of isotropic turbulence. Since most of the energy in vorticity generated by these grains is in far-subcritical shear waves with inclination $\theta<10^{\circ}$, critical angle effects are negligible for the two proceeding shocks. This could change however, if there was an additional shock to further refract these modes to near-subcritical inclinations. Suppose the turbulence is subjected to an initial shock of $m_1=2.2$, and the strengths of two proceeding shocks are chosen as $m_2$ and $m_3$.  For the axisymmetric turbulence from the grain simulation, a maximal difference of factor $R_{\rm max}=1.128$ is possible upon swapping two shocks of strengths $m_2=1.54$ and $m_3=4$. This is a greater difference than for initially isotropic turbulence in the two-shock problem and also in this three-shock problem where $R_{\rm max}=1.110$.

Both the NIC and IC limiting cases we studied in Section~\ref{sec:results} are only applicable when the time separation between shocks is sufficiently shorter or longer than the timescale over which turbulence returns to isotropy after each shock. Direct numerical simulations have shown the vorticity returns to isotropy after ten convected Kolmogorov timescales, $\tau_{\eta} = \sqrt{\nu / \epsilon}$, where $\nu$ is the kinematic viscosity and $\epsilon$ the dissipation rate of turbulent kinetic energy \cite{larsson2013}. However, the turbulent velocity was found to remain anisotropic in their simulated domains, and one might expect a return to isotropy over a longer distance away from the shock. Provided this distance is greater than the necessary shock separation to avoid overtaking of shocks, the first limiting case can be achieved with minimum shock separation. The second case is possible by making the shock separations longer than both the required shock separation and the distance needed for return to isotropy; with the caveat that if the separations are too great, non-negligible viscous dissipation and non-linear effects can occur that we have neglected in this model.

However, even for situations in between the two cases, where the turbulence begins an incomplete return to isotropy before the next shock, one may anticipate results to be a mix of the two limiting cases. For example, shock ordering does not matter for the isotropized case, while TKE amplification in the NIC limit is sensitive to it. One may expect for realistic regimes in between these two cases, that shock ordering differences in TKE amplification will still manifest, but to a lesser degree.

To model such scenarios, where turbulence evolution is important, is outside the scope of this work. Such evolution may be governed by nonlinear processes not included in LIA \cite{larsson2009} and require the inclusion of viscous dissipation.

Further, any caveats of LIA will also appear in our model, even for the two limiting cases. We've assumed weak turbulence in order to use LIA, and to treat the jumps in background quantities as unperturbed from classical Rankine-Hugoniot shock-jump relations for density, temperature and mean flow speed. Under continuous amplification by many shocks, the turbulence could eventually become energetic enough relative to the shocks, that this assumption is broken. In DNS, disagreement with LIA in TKE amplification and mean Rankine-Hugoniot jumps is found for higher turbulent Mach numbers. However, there have been more recent theories that treat finite Mach turbulence and the modifications to jumps in background quantities \cite{chen2019}, that could also be adapted similarly for our model here. Moreover, the velocity anisotropies in DNS are observed to be different from the predictions of LIA \cite{larsson2009}, which would alter the NIC limit.

Here we have only taken the vortical contributions to turbulence in calculating amplifications. However, there is also an acoustic pressure contribution that we have neglected. For most shock strengths, this acoustic contribution is of a few percent relative to that of the vortical amplification, but can be dominant in the weak shock limit \cite{wouchuk2009}. Thus our calculation of the weak shock limit should be modified to treat the acoustic portion too. In addition, our work neglects the effects of density non-uniformity, which has been shown to be important for turbulence generation by shocks in certain contexts such as the aforementioned examples of grain structure in ICF ablators \citep{davidovits2022b, li2024}. However, thermal conduction in such cases can smooth the post-shock density field, so that any subsequent shocks following the first would primarily interact with the vorticity field.

\section{Conclusion}
\label{sec:conclusion}

We use a linear analytical model following the single-shock results (LIA) of Ribner \cite{ribner1954} to study the interaction of an incompressible, isotropic spectrum of turbulence with multiple planar shocks in two limits: 1) The turbulence fully returns to isotropy between each shock (IC limit), or 2) Shocks occur rapidly in time such that the turbulence spectrum does not evolve between shocks (NIC limit). We assume the jumps in background quantities are unperturbed for sufficiently weak turbulence. For a variety of number and strengths of shocks, we examine the amplification of vortical turbulent kinetic energy (TKE) across multiple shocks, opting to neglect the acoustic contribution. By extracting the effective polytropic indices for the turbulent pressure amplification, we infer general properties of the multi-shock compression of turbulence. To further inform the sensitivity to shock ordering in the NIC limit, we also consider the multi-shock interaction with single-mode shear waves for certain example inclinations with respect to the initial shock.

We find that choosing to launch a greater number of weaker shocks to reach a fixed final background state will generally result in greater amplification of turbulence than a series of fewer stronger shocks. In particular, in the IC limit, we find maximal TKE amplification (without regard to relative temperature amplification) for a series of weak shocks with strength $m\approx 1.1$. As shock strength tends towards unity for this IC limit, we infer the polytropic index to be $5/3$, revealing that an arbitrarily weak multi-shock compression of turbulent pressure is equivalent to that of an isotropic, adiabatic volumetric compression. In the NIC limit of many weak shocks, the polytropic index is instead super-adiabatic, for example $n\approx1.83$ for $m=1.1$. As the multi-shock amplification of turbulence becomes less efficient towards greater strength shocks, the associated polytropic index decreases for both IC and NIC limits, with the latter becoming adiabatic, $n=5/3$, at $m=1.3$. The polytropic index of the temperature instead increases with shock strength, and so stronger shocks convert a greater proportion of energy into heating than into compression and turbulent amplification.

We show how the anisotropy in the NIC limit affects the multi-shock amplification of turbulence, both in its components relative to the shock fronts and overall. Stronger shocks magnify the anisotropy, accelerating the departure of the NIC limit from the IC limit by reducing TKE amplification and its polytropic index with respect to multi-shock compression, as the number of shocks increases. Therefore, the polytropic index associated with the NIC limit is generally dependent on the shock history.

In addition for the NIC limit, TKE amplification is sensitive to not only the choice of shock strengths, but the order in which each shock is launched. With two shocks, for example, differences in amplification of up to $8.4\%$ are possible for initially isotropic turbulence, when the two shocks are switched in order. Single-mode shear waves can exhibit far greater sensitivity to ordering when the shear wave inclination is within a band below the critical angle, with up to $219\%$ amplification difference upon swapping two shocks of strengths $m=1.94$ and $m=3.54$ for an example initial inclination of $15^\circ$. It is therefore likely that certain anisotropic spectra of turbulence can be highly sensitive to the ordering of certain strength shocks.

Our use of LIA to examine multi-shock compression of turbulence in two limits informs a simple intuition for how turbulence may behave under multi-shock compression and the impact of shock strengths. We reveal an interesting sensitivity of TKE amplification to the order in a choice of strengths of shocks. While one could intuit behavior in a regime between the two limits, further work on a model that includes the evolution of the turbulence spectrum in between shocks, with viscous dissipation, remains to be explored.

\begin{acknowledgments}

This work benefited from useful conversations with Christopher Weber. M.F.Z.~and N.J.F.~acknowledge support, in part, from DOE DE-AC02-09CH1146 and by the Center for Magnetic Acceleration, Compression, and Heating (MACH), part of the U.S.~DOE-NNSA Stewardship Science Academic Alliances Program under Cooperative Agreement No.~DE-NA0004148. M.F.Z.~acknowledges support from the Defense Science and Technology Internship program at LLNL. This work was performed under the auspices of the U.S.~Department of Energy by the Lawrence Livermore National Laboratory under Contract No.~DE-AC52-07NA27344. S.D.~was supported by the LLNL-LDRD Program under Project No.~20-ERD-058.

\end{acknowledgments}


\end{document}